\documentstyle[preprint,aps,graphicx]{revtex}

\def\red{
}
\def\black{
}

\def\URLtilde{\lower0.2em\hbox{$\tilde{\phantom{a}}$}}
\def\mycomm#1{\hfill\break\strut\kern-3em{\red\tt ====> #1\black}\hfill\break}
\def\mycommNL#1{\strut\kern-3em{\tt ====> #1}\hfill\break}

\catcode`\@=11 % This allows us to modify PLAIN macros.
\def\lsim{\mathrel{\mathpalette\@versim<}}
\def\gsim{\mathrel{\mathpalette\@versim>}}
\def\@versim#1#2{\vcenter{\offinterlineskip
             \ialign{$\m@th#1\hfil##\hfil$\crcr#2\crcr\sim\crcr } }}
\catcode`\@=12 % at signs are no longer letters

\def\eqref#1{(\ref{#1})}

\makeatletter
\def\hlinewd#1{\noalign{\ifnum0=`}\fi
\hrule \@height #1 \futurelet \reserved@a\@xhline}
\def\hwhiteline{\noalign
{\ifnum0=`}\fi\hrule
\@height 0pt\vskip 1.0ex\futurelet \reserved@a\@xhline}
\makeatother
\def\gray{\special{ps: 0.40 setgray}}
\def\black{\special{ps: 0.0 setgray}}

\newcommand{\mydraft}{
\newcount\timecount
\newcount\hours \newcount\minutes \newcount\temp \newcount\pmhours

\hours = \time
\divide\hours by 60
\temp = \hours
\multiply\temp by 60
\minutes = \time
\advance\minutes by -\temp
\def\hour{\the\hours}
\def\minute{\ifnum\minutes<10 0\the\minutes
         \else\the\minutes\fi}
\def\clock{
\ifnum\hours=0 12:\minute\ AM
\else\ifnum\hours<12 \hour:\minute\ AM
\else\ifnum\hours=12 12:\minute\ PM
         \else\ifnum\hours>12
          \pmhours=\hours
          \advance\pmhours by -12
          \the\pmhours:\minute\ PM
          \fi
         \fi
\fi
\fi
}
\def\fullclock{\hour:\minute}
\begin{centering}
\gray
\font\Hugett =cmtt12 scaled\magstep4
\hbox{\Hugett Draft:\today,\clock}
\black
\end{centering}
\vskip -1.7cm
$\phantom{a}$
} % end of \draft definition
\def\beq#1{\begin{equation} \label{#1}}
\def\eeq{\end{equation}}

\def\ket#1{\left\vert #1\right\rangle}

\newskip\humongous \humongous=0pt plus 1000pt minus 1000pt

\newif\ifdtup

\def\3s{\hbox{\small $\bar{\bf s}\bar{\bf 3}$}}
\def\6s{\hbox{\small $\bar{\bf s}{\bf 6}$}}

\begin{document}
{\tighten
     \preprint {\vbox{
     \hbox{$\phantom{aaa}$}
     \vskip-0.5cm
%\hbox{\today}
%\hbox{}
\hbox{TAUP 2827/06}
\hbox{WIS/06/06-JULY-DPP}
\hbox{ANL-HEP-PR-06-52}
}}

\title{A tetraquark model for the new $X(1576)$ $K^+K^-$ resonance}
\author{Marek Karliner\,$^{a}$\thanks{e-mail: \tt marek@proton.tau.ac.il}
\\
and
\\
Harry J. Lipkin\,$^{a,b}$\thanks{e-mail: \tt
ftlipkin@weizmann.ac.il} }
\address{ \vbox{\vskip 0.truecm}
$^a\;$School of Physics and Astronomy \\
Raymond and Beverly Sackler Faculty of Exact Sciences \\
Tel Aviv University, Tel Aviv, Israel\\
\vbox{\vskip 0.0truecm}
$^b\;$Department of Particle Physics \\
Weizmann Institute of Science, Rehovot 76100, Israel \\
and\\
High Energy Physics Division, Argonne National Laboratory \\
Argonne, IL 60439-4815, USA\\
}
\maketitle
%\mydraft

\begin{abstract}%
\strut\vskip-1.0cm
We discuss the likely tetraquark interpretation of the 
$X(1576)$ $K^+K^-$ isovector resonance recently reported by BES
with $J^{PC}=1^{{-}{-}}$.
We point out that if this interpretation as a four-quark state 
is correct,
the $s u \bar s \bar d$ tetraquark decays might have striking signatures.
We also provide predictions for possible analogous tetraquarks 
involving heavy quarks -- $c q \bar c \bar q$ and $b q \bar b \bar q$.

\end{abstract}% %}
% end tighten

\vfill\eject

\section{The new isovector $K^+K^-$ resonance}

We have noticed with interest the new $K^+K^-$
resonance\cite{beskk} with $J^{PC}=1^{--}$ , and a pole mass at
$1576^{{+}49}_{{-}55}{\rm (stat)}^{{+}98}_{{-}91}{\rm (syst)}
- i\left( 409^{+11}_{{-}12}{\rm (stat)}^{+32}_{{-}67}{\rm
(syst)} \right)$ MeV which seems to be a four-quark state.
Since it is produced with a pion from the isoscalar $J/\psi$, which
has odd $G$-parity it must be isovector and have even $G$-parity.

Since no nonstrange even-G vector resonances like
$\rho-\rho$ or $\pi-\pi$ have been reported in this mass
range the new $K^+K^-$ resonance seems to contain a strange
quark pair. An isovector particle containing a strange
quark pair must necessarily contain an additional isovector
nonstrange pair to make an isovector.  Thus this new state
must contain a minimum of two quark-antiquark pairs.
It is also worth noting that this 
state cannot be a $\bar s s$ hybrid, 
since a state of a strange quark pair 
and gluons is isoscalar and cannot be isovector.

A recent examination of four-quark (tetraquark)
states\cite{NewTetra} has emphasized the states including
heavy quarks near the two-meson threshold may well exist as
bound tetraquarks. Applying this approach to the two-kaon
system gives two diquark-antidiquark configurations with
masses above the two-kaon threshold with ratios of masses
to the mass of two kaons of 1.21 and 1.16 respectively.
These results are for $S$-wave systems and neglect spin
dependence.
They are therefore lower than the
mass of the new state which must have a $P$-wave to
explain the negative
parity. They are therefore in a reasonable ball park,
since experimentally
$m_X/(2 m_K) \approx 1.6$.

In contrast with the $1^{--}$ state reported by BES
\cite{beskk},
the $S$-wave tetraquark with these flavor quantum numbers
probably breaks up so fast that it is too broad to be
seen.

At this stage any further calculation of orbital and spin
effects will be
highly model dependent and unreliable. However, the
diquark-antidiquark model
\cite{NewTetra} makes clear flavor predictions that are
easily tested in
experiment.
We consider here predictions relating the decays of the
new $X(1576)$ resonance. Although only $X^0$ has been observed
experimentally,
existence of $X^+$ and $X^-$ follows from $X$ having $I=1$. Since
decays involving charged pions in the final state are much easier to
observe, we focus on these:

We first note that the dominant decay mode of a tetraquark 
is the ``fall-apart mode" in which the two quarks 
and two antiquarks rearrange into two mesons and
separate. No new quark-antiquark pairs are created or destroyed. This immediately leads to a
selection rule for a diquark($us$)-antidiquark($\bar d\bar s$) model for
the new $X(1576)$ resonance. 

\beq{selrule}
BR(X^+ \rightarrow \pi^+ \pi^o)= 0
\eeq

This selection rule provides the crucial distinction between tetraquark and
$q \bar q$ models. A positively
charged $q \bar q$ must be in an octet of $SU(3)_{flavor}$. The $\pi \pi$
 decay amplitude is required by $SU(3)$ symmetry to be 
proportional to the same reduced matrix element as the $K \bar K$ and cannot
vanish for a hadron decaying into $K \bar K$. 

A further prediction for the tetraquark model is obtained by noting that 
there are two ways in which a $us$ diquark and a $\bar d
\bar s$ antidiquark can rearrange to make two mesons
\beq{twoways}
\ket{us,\bar d \bar s}
\rightarrow \ket{M(u\bar d),M (s\bar s)}; ~ ~ ~
\ket{us;\bar d \bar s}
\rightarrow \ket{M(u\bar s),M (s\bar d)}
\eeq

These two decay modes must be equal in the flavor-$SU(3)$ symmetry
limit since there is no
preferred way that a
quark in the diquark can choose a particular antiquark in
the antidiquark.

$SU(3)_f$ breaking will make the two modes somewhat different, because the 
color-magnetic spin-dependent interaction depends on the quark masses, but
we expect the approximate equality to hold at least as well as most
$SU(3)_f$ relations.

The approximate equality of the two modes \eqref{twoways}
immediately gives predictions relating the decays of the
new $X(1576)$ resonance.

\beq{predpv}
\tilde \Gamma (X^+ \rightarrow K^+ \bar K^{*o})=
\tilde \Gamma (X^+ \rightarrow K^{*+} \bar K^o)
\approx
\tilde \Gamma (X^+ \rightarrow \pi^+ \phi) 
\approx
\tilde \Gamma (X^+ \rightarrow \rho^+ \eta) +
\tilde \Gamma (X^+ \rightarrow \rho^+ \eta')
\eeq
where $\tilde \Gamma $ denotes the partial width neglecting
phase space
corrections, and the conventional nonet description is used for
the
pseudoscalar mesons with $SU(3)$ flavor symmetry.

Experimental confirmation of the selection rule (\ref{selrule}) and the
predictions (\ref{predpv}) would provide
unambiguous support for the tetraquark description, since other models relate
the observed $K\bar K$ decay very differently to these other 
two-pseudoscalar decays.

The first equality in eq.~\eqref{predpv} is  model
independent and follows from $G$-parity. The remaining equalities have
symmetry breaking corrections which can be averaged out in some
approximation by rewriting this relation as

\beq{predpvav}
\tilde \Gamma (X^+ \rightarrow K^+ \bar K^{*o}) +
\tilde \Gamma (X^+ \rightarrow K^{*+} \bar K^o)
\approx
\tilde \Gamma (X^+ \rightarrow \pi^+ \phi)  +
\tilde \Gamma (X^+ \rightarrow \rho^+ \eta) +
\tilde \Gamma (X^+ \rightarrow \rho^+ \eta')
\eeq

The prediction (\ref{predpv}) relates OZI-allowed decays to OZI-forbidden
decays. Although the equality may not be exact, because of flavor symmetry
breaking, there is clearly no OZI suppression predicted.

Further support for a tetraquark description of the  
$X(1576)$ would be a strong $\phi\pi$ branching ratio, since the $\phi\pi$
decay is OZI-forbidden for a normal quark-antiquark
resonance. 

Recent data on the decay $J/\psi \rightarrow \phi \pi \pi$ show
no $\phi$-$\pi$ resonance near the new $X(1576)$ resonance\cite{Zou}. If this 
leads to a strong disagreement
with the prediction \eqref{predpvav}, it will pose a serious difficulty
for most models. A model containing an isovector $q\bar q$ pair and no
additional quarks must be in a flavor $SU(3)$ octet which requires 
comparable $K^+K^-$ and $\pi^+\pi^-$ decays. A model which contains both
an isovector pair and a strange quark pair can decay by quark 
rearrangement into $\phi \pi$. There is no known OZI rule for tetraquarks
to suppress this decay. 

\section{OZI Violation as the key to multiquark states}

Violation of the OZI rule provides a useful signature to distinguish
between
decays of normal mesons and tetraquarks. In the two ways
(\ref{twoways}) in
which a $us$ diquark and a $\bar d \bar s$ antidiquark can rearrange
to make
two mesons both ways have the same topology and the two are
equivalent. When
these final states are produced by the decay of a normal $q\bar q$
isovector
meson, the transition to the state
$\ket{M(u\bar d),M (s\bar s)}$ is forbidden by the OZI rule
while the transition to the state $\ket{M(u\bar s),M (s\bar d)}$ is allowed.

This flavor-exchange symmetry can thus be used as a test of tetraquark
models for other anomalous quarkonium states including heavy quarks.  All
isovector states that decay into heavy quark pairs are generally
recognizable as multiquark states, since such a decay of a nonstrange isovector
state by the creation of a heavy $\bar qq$ is expected to be suppressed.
The analogues of eq. (\ref{predpv}) for charm and bottom are:

\beq{predpvc}
\tilde \Gamma (X_c^+ \rightarrow D^+ \bar D^{*o})=
\tilde \Gamma (X_c^+ \rightarrow D^{*+} \bar D^o)
 \approx
\tilde \Gamma (X_c^+ \rightarrow \pi^+ J/\psi) 
 \approx
\tilde \Gamma (X_c^+ \rightarrow \rho^+ \eta_c)
\eeq
\beq{predpvb}
\tilde \Gamma (X_b^+ \rightarrow B^+ \bar B^{*o})=
\tilde \Gamma (X_b^+ \rightarrow B^{*+} \bar B^o)
\approx
\tilde \Gamma (X_b^+ \rightarrow \pi^+ \,\Upsilon\,) 
\approx
\tilde \Gamma (X_b^+ \rightarrow \rho^+ \eta_b)
\eeq
In analogy with eq.~(\ref{predpv})
the first equalities in eqs.~\eqref{predpvc} and \eqref{predpvb} are 
model 
independent and follow from $G$-parity. The remaining equalities have 
symmetry breaking corrections which can be averaged out in some 
approximation by rewriting these relations as

\beq{predpvcav}
\tilde \Gamma (X_c^+ \rightarrow D^+ \bar D^{*o}) +
\tilde \Gamma (X_c^+ \rightarrow D^{*+} \bar D^o)
=
\tilde \Gamma (X_c^+ \rightarrow \pi^+ J/\psi)  +
   \tilde \Gamma (X_c^+ \rightarrow \rho^+ \eta_c)
\eeq
\beq{predpvbav}
\tilde \Gamma (X_b^+ \rightarrow B^+ \bar B^{*o}) +
\tilde \Gamma (X_b^+ \rightarrow B^{*+} \bar B^o)
=
\tilde \Gamma (X_b^+ \rightarrow \pi^+ \Upsilon) +
\tilde \Gamma (X_b^+ \rightarrow \rho^+ \eta_b)
\eeq
One should keep in mind, however,
that even with this averaging 
relations \eqref{predpvcav} and \eqref{predpvbav} 
receive significant corrections due to $m_b$, $m_c \gg m_u, m_d$.

This reasoning can be extended to
identify isoscalar heavy-quark tetraquark states. We note that the
flavor-exchange symmetry relates the two ways in which a $dQ$ diquark and a
$\bar d \bar Q$ antidiquark can rearrange to make two mesons, even though one
is OZI allowed and one is OZI forbidden for the decay of a light quark meson.

\beq{twowaysQ} \ket{dQ,\bar d \bar Q}
\rightarrow \ket{M(d\bar d),M (Q\bar Q)}; ~ ~ ~ \ket{dQ;\bar d \bar Q}
\rightarrow \ket{M(d\bar Q),M (Q\bar d)}
\eeq

This immediately gives predictions identifying possible anomalous isoscalar
strangeoniumm, charmonium and bottomonium states with even parity and even 
charge conjugation,
denoted as $X_s^{(I=0)}$, $X_c^{(I=0)}$ and $X_b^{(I=0)}$, as cryptoexotic multiquark 
states by the
absence of OZI suppression:

\beq{predppc}
2\,\tilde \Gamma (X_c^{(I=0)} \rightarrow D^+ \bar D^-)
\approx
\tilde \Gamma (X_c^{(I=0)} \rightarrow \eta\, \eta_c) +
\tilde \Gamma (X_c^{(I=0)} \rightarrow \eta'\, \eta_c)
\eeq

\beq{predppb}
2\,\tilde \Gamma (X_b^{(I=0)} \rightarrow B^+ \bar B^-)
\approx
\tilde \Gamma (X_b^{(I=0)} \rightarrow \eta \,\eta_b) +
\tilde \Gamma (X_b^{(I=0)} \rightarrow \eta'\, \eta_b)
\eeq

\beq{predvvs}
2\,\tilde \Gamma (X_c^{(I=0)} \rightarrow K^{*+} \bar K^{*-})
\approx
\tilde \Gamma (X_c^{(I=0)} \rightarrow \phi \,\omega)
\eeq

\beq{predvvc}
2\,\tilde \Gamma (X_c^{(I=0)} \rightarrow D^{*+} \bar D^{*-})
\approx
\tilde \Gamma (X_c^{(I=0)} \rightarrow J/\psi \,\omega)
\eeq

\beq{predvvb}
2\,\tilde \Gamma (X_b^{(I=0)} \rightarrow B^{*+} \bar B^{*-})
\approx
\tilde \Gamma (X_b^{(I=0)} \rightarrow \Upsilon\,\omega)
\eeq

The additional factor of 2 on the l.h.s. of 
eqs.~\eqref{predppc}--\eqref{predvvb} 
results from the that $X_Q^{(I=0)}$ has both a
$d Q \bar d \bar Q$ and a $u Q \bar u \bar Q$ component, so 
that eq.~\eqref{twowaysQ}
  should have a companion equation with $d$ replaced by $u$.
Both of these amplitudes contribute to the right hand side of 
eqs.~\eqref{predppc}--\eqref{predvvb}.
Only one contributes to the left hand side. The other one contributes to
the final state with isospin partners of the $q \bar Q$ mesons, such
as $D^o \bar D^o$, etc.

The direct analog of eqs.~\eqref{predppc} and \eqref{predppb} 
for strangeonium decays 
to the $\eta$ and $\eta'$  are more complicated
because of the $\eta-\eta'$ mixing which does not exist for $\eta_c$ and
$\eta_b$. For the transitions to the strange and nonstrange pseudoscalar mesons 
denoted by $\eta_s$ and $\eta_n$ without flavor mixing, we have

\beq{predpps}
\tilde \Gamma (X_s^{(I=0)} \rightarrow K^+ \bar K^-)=
\tilde \Gamma (X_s^{(I=0)} \rightarrow \eta_n \,\eta_s)
\eeq

\section{Alternative tetraquark model - a $K\bar K$ molecule}

If there is a strong disagreement
with the prediction \eqref{predpvav}, it will pose a serious difficulty
for most models. 

    One model that might explain the explain an absence of a $\phi \pi$ decay
mode or a violation of the prediction \eqref{predpvav} is a $K^+K^-$ molecule
where the two kaons are sufficiently far apart so that they cannot exchange
quarks. However, if the $K^+$ and $K^-$ are that far apart they are very
likely  also too far apart to annihilate a $u\bar u$ pair and create a $d\bar
d$ pair to make a $K^o\bar K^o$ molecule. In this case the $X$ is not an
isospin eigenstate but is a pure $K^+K^-$ molecule. This is an isospin mixture
with equal isovector and isoscalar components. It should be  produced in
$J/\psi$ decays not only with the isovector pion but also via its isoscalar
component with the isoscalar $\eta$ and $\eta'$ or the isoscalar $\omega$.
These final states should also be seen if the $X$ is indeed an isovector but
has an isoscalar partner, analogous to the $\rho$ and $\omega$ doublet.

\section{Conclusions}

We present experimental predictions for decays of tetraquarks that 
can distinguish between tetraquark and other models for hadronic states.
These predictions provide a  serious test for the new $X(1576)$ resonance.
An isovector $K^+K^-$ resonance which does not decay into $\pi^+\pi^-$ 
cannot be a flavor $SU(3)$ octet and must contain a strange quark pair in 
addition to a nonstrange isovector pair. If its decays also violate the
tetraquark predictions, it must have a new multiquark structure like a molecule.

\section*{Acknowledgements}

We thank  Bingsong Zou for correspondence on the BES data and for pointing
out an error in one of the channels considered in the previous version of
this paper.
The research of M.K. was supported in part by a grant from the
Israel Science Foundation administered by the Israel
Academy of Sciences and Humanities.
The research of H.J.L. was supported in part by the U.S. Department
of Energy, Division of High Energy Physics, Contract W-31-109-ENG-38.

%----------------------------------------------------------------------
% This prevents REFERENCES from forcing a page break
%\def\newpage{\vskip10ex}
%
\catcode`\@=11 % This allows us to modify PLAIN macros
\def\references{
\ifpreprintsty \vskip 10ex
%\ifpreprintsty \newpage
%
\hbox to\hsize{\hss \large \refname \hss }\else
\vskip 24pt \hrule width\hsize \relax \vskip 1.6cm \fi \list
{\@biblabel {\arabic {enumiv}}}
{\labelwidth \WidestRefLabelThusFar \labelsep 4pt \leftmargin \labelwidth
\advance \leftmargin \labelsep \ifdim \baselinestretch pt>1 pt
\parsep 4pt\relax \else \parsep 0pt\relax \fi \itemsep \parsep \usecounter
{enumiv}\let \p@enumiv \@empty \def \theenumiv {\arabic {enumiv}}}
\let \newblock \relax \sloppy
     \clubpenalty 4000\widowpenalty 4000 \sfcode `\.=1000\relax \ifpreprintsty
\else \small \fi}
\catcode`\@=12 % at signs are no longer letters
%-----------------------------------------------------------------
%{\tighten

 % end of global \tighten
\end{document}